\pdfoutput=1
\documentclass{article}

\usepackage{microtype}
\usepackage{graphicx}
\usepackage{subfigure}
\usepackage{booktabs}
\usepackage{xcolor}
\usepackage{hyperref}

\usepackage[accepted]{mlsys2025}
\makeatletter
\renewcommand{\printAffiliationsAndNotice}[1]{%
\stepcounter{@affiliationcounter}%
{\let\thefootnote\relax\footnotetext{\hspace*{-\footnotesep}#1%
\forloop{@affilnum}{1}{\value{@affilnum} < \value{@affiliationcounter}}{%
\textsuperscript{\arabic{@affilnum}}\csname @affilname\the@affilnum\endcsname}.}}}
\makeatother

\usepackage{amsmath}
\usepackage{amssymb}
\usepackage{xspace}
\usepackage{multirow}
\usepackage[capitalise,nameinlink]{cleveref}
\renewcommand{\topfraction}{0.9}
\renewcommand{\bottomfraction}{0.8}
\renewcommand{\textfraction}{0.07}

\usepackage{enumitem}
\setlist{nosep}

\newif\ifshowplans
\showplansfalse
\definecolor{notecolor}{RGB}{190,90,0}

\definecolor{nearfinal}{RGB}{150,20,40}

\ifshowplans
  \newcommand{\note}[1]{{\color{notecolor}\sffamily\footnotesize[Note: #1]}}
  \newcommand{\verify}[1]{{\color{notecolor}\sffamily\footnotesize[Check: #1]}}
  \newcommand{\citeneeded}[1]{{\color{red}\sffamily\footnotesize[cite: #1]}}
\else
  \newcommand{\note}[1]{}
  \newcommand{\verify}[1]{}
  \newcommand{\citeneeded}[1]{}
\fi

\providecommand{\AtAppendixStart}{}

\mlsystitlerunning{Retraining-Free Accurate Inference in Analog CIM Accelerators}

\hypersetup{pdftitle={Enabling Retraining-Free Accurate Inference in Analog Compute in Memory Accelerators},pdfauthor={George Karfakis, Puneet Gupta}}
\begin{document}

\twocolumn[
\mlsystitle{Enabling Retraining-Free Accurate Inference in Analog Compute in Memory Accelerators}

\begin{mlsysauthorlist}
\mlsysauthor{George Karfakis}{ucla}
\mlsysauthor{Puneet Gupta}{ucla}
\end{mlsysauthorlist}

\mlsysaffiliation{ucla}{Department of Electrical and Computer Engineering, University of California, Los Angeles, CA, USA}

\mlsyskeywords{}

\vskip 0.3in

\begin{abstract}
Analog compute-in-memory (CIM) accelerators compute matrix-vector products directly inside memory arrays,
eliminating weight movement and promising order-of-magnitude gains in inference efficiency. However, converter
and device errors make pretrained models unusable on these arrays, and hardware-aware retraining is costly. We
enable retraining-free, accurate inference of pretrained models on simple analog arrays with low-cost, shared
calibration. Our key observation is that analog error depends on how each layer is
represented on the array, and that this representation can be changed offline without changing the computation.
We center the inputs and apply Hadamard rotations designed for analog arrays, which keep shared reference errors
spread and confine input mixing to each array. Across CNNs, vision transformers and LLMs up to 14B parameters on
PCM and ReRAM, the method brings vision models and larger LLMs within a few points of floating-point accuracy and matches or
approaches published hardware-aware training without training any weights. On ReRAM with 6-bit ADCs, it restores ViT-B/16 from 37.4\% to
82.9\% and Phi-4 from 27.0\% to 72.2\%, both within 2.2 points of floating-point accuracy.
\end{abstract}
]

\printAffiliationsAndNotice{}

\section{Introduction}
\label{sec:intro}

\label{sec:motivation}

Moving model weights between memory and compute can cost more time and energy than the arithmetic itself,
especially at the small batch sizes of latency-critical and edge inference \cite{horowitz2014energy, Gholami2024MemoryWall}.
Analog compute-in-memory (CIM) sidesteps this issue by exploiting analog effects to perform matrix-vector operations directly on memory arrays. \cite{Sebastian2020, Verma2019IMC}. Dense non-volatile memory
(NVM) arrays can hold all of a model's weights throughout inference, so no weights stream from external
memory \cite{Sebastian2020, Burr2023}. For models that fit, system studies project 40 to 140 times the
energy efficiency of an NVIDIA A100 \cite{Jain2023Mesh}.

These efficiency gains come with analog errors that can substantially reduce model accuracy. Precise analog-to-digital converters (ADCs) can be very costly in area and energy, so designs tend to use low-precision ADCs that
quantize and clip the array outputs \cite{Shafiee2016ISAAC, Murmann2024ADC, Xiao2022}. NVM devices introduce additional errors through programming noise, read noise and drift of the stored weights over time
\cite{Joshi2020, Nandakumar2019, Mackin2022, He2021Relaxation}. These errors can make pretrained
models unusable: in our ReRAM simulation with 6-bit ADCs, one day after programming, ViT-B/16
falls from 84.5\% to 37.4\%, and Phi-4 from 74.4\% to 27.0\% accuracy
over studied benchmarks.

A widely used approach to recovering this accuracy is hardware-aware (HWA) training, which injects analog errors into the forward pass so that the model learns to tolerate them \cite{Joshi2020, Kariyappa2021, Rasch2023,  Chen2025, Zhou2025,
Buchel2025AFM, Li2023ApproxTrain}. However, this approach  requires access to training data and can be very costly, becoming a serious barrier to analog deployment. The analog foundation models of \citet{Buchel2025AFM}, for instance, train on billions of tokens.

\citet{Buchel2025AFM} explicitly call for a post-training method that recovers accuracy under
analog errors, but find that the digital post-training quantization (PTQ) methods provide
insufficient robustness and lose accuracy when input ranges are held fixed. We seek to recover accuracy while
retaining shared, static array settings, which allow reference and control circuitry to be reused across
columns and simplify calibration \cite{KhaddamAljameh2022HERMES,Ambrogio2023,Yang2026PerColumn}. We develop methods so that
off-the-shelf pretrained models tolerate both converter and device errors under these settings {\em without
retraining or modifying the arrays}.

\label{sec:overview}

We observe that the representation of an otherwise identical neural computation determines several distinct physical error mechanisms in analog CIM, and each mechanism can be attacked by a corresponding function-preserving representation change. We first center the computation to remove static, precomputable work from the analog path. We then apply computationally invariant  isometric transformations to express the remaining computation in a basis better suited to the array's noisy conductances and
low-precision partial sums. Table~\ref{tab:error-map} summarizes how the array's error mechanisms guide the choice and
construction of these transformations.

\begin{table*}[t]
\centering
\footnotesize
\setlength{\tabcolsep}{4pt}
\caption{How the structure of the computation and the array's constraints determine our isometric
reparameterization.}
\label{tab:error-map}
\begin{tabular}{@{}p{3.0cm}p{6.2cm}p{5.6cm}l@{}}
\toprule
Observation & Array effect & Our fix & Sec. \\
\midrule
Static input components & Static work picks up converter and device error on every input & Center inputs and restore digitally & \ref{sec:m-static} \\
Concentrated partial outputs & Peaks force a clip--quantization tradeoff under one ADC range & Regular output rotation to lower peaks & \ref{sec:m-adc} \\
Concentrated weights & Outliers set the mapping scale and amplify conductance perturbations & Input and output rotations to reduce weight peaks & \ref{sec:m-device} \\
\midrule
\emph{Constraint:} shared/common-mode terms & Plain Hadamard can concentrate common-mode (e.g., reference) error & Regular Hadamard blocks ($R\mathbf{1}=\mathbf{1}$) & \ref{sec:m-adc} \\
\emph{Constraint:} tiled conversion & Cross-tile mixing changes per-tile quantized partial sums & Tile-local input rotations & \ref{sec:intertile} \\
\bottomrule
\end{tabular}
\end{table*}

We reparameterize the weights offline and apply matched digital transforms during inference, without
retraining or changing the array hardware. In
the ReRAM setting above, they raise ViT-B/16 and Phi-4 accuracy to 82.9\% and 72.2\%, within 2.2
points of floating-point execution. These gains require neither retraining nor dynamic input ranges: each array uses one static weight
mapping scale, input range, and output range.

Our contributions are:
\begin{itemize}
  \item We analyze how analog CIM error depends on the representation of weights and activations.
 \item We propose a retraining-free reparameterization (centering + regular output rotations + tile-local input rotations)
    for robust inference on unmodified arrays with static calibration.
 \item We evaluate CNNs, vision transformers (ViTs) and large language models (LLMs) on PCM and ReRAM with
    4- to 8-bit ADCs, showing consistent accuracy recovery and gains over strong baselines.
\end{itemize}

\section{Background and Related Work}
\label{sec:background}

\subsection{Analog CIM systems}
\label{sec:arrays}

\begin{figure}[t]
 \centering
 \includegraphics[width=\columnwidth]{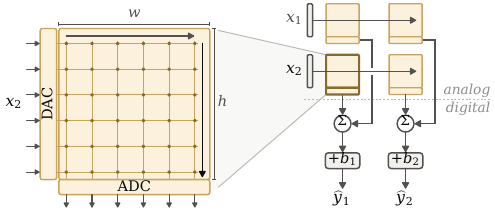}
 \caption{An analog CIM array of width $w$ and height $h$ with its DAC and ADC (left), and a weight matrix
split into tiles whose partial sums are added digitally before the bias (right).}
 \label{fig:tiling}
\end{figure}

An analog CIM array computes a matrix-vector product using the memory cells that store its weights.
Each device produces a current proportional to the product of its stored weight and input voltage on
its row. The sum of currents along each column form the output \cite{Sebastian2020, LeGallo2023}, which
an ADC reads. Larger matrices are tiled across arrays (Figure~\ref{fig:tiling}). Each array's partial
output is converted separately, and digital logic combines the converted partial sums
\cite{Shafiee2016ISAAC, KhaddamAljameh2022HERMES}.

The assignment of weight tiles to arrays distinguishes two dataflows. Partially weight-stationary (PWS)
designs reuse arrays for successive tiles, reloading weights for every tile. SRAM favors this dataflow
\cite{Verma2019IMC}. Fully weight-stationary (FWS) designs instead keep each tile in its own array, avoiding reloading resident
model-weight tiles during inference. NVM-based CIM arrays favor FWS operation due to their high
density and efficiency \cite{LeGallo2023, Ambrogio2023, KhaddamAljameh2022HERMES}. We focus on FWS NVM arrays.

We only target static-weight layers mapped to CIM. Dynamic operations, such as the matrix
products in attention, use operands generated at runtime, cannot exploit fixed weight residency and run digitally
\cite{Wolters2024CIM, Burr2023, Karfakis2026MXFormer, Spoon2021}.

\subsection{Analog compute error sources}
\label{sec:mvm}

Analog matrix-vector products accumulate errors in three distinct stages of the compute path: converting the inputs into analog signals using a digital-to-analog converter (DAC), computing with the stored conductances, and digitizing the outputs with an ADC.

\textbf{Inputs.} A DAC encodes each input activation as an analog signal
(voltage, current or pulse duration) that drives one row of the array. Its finite resolution
over a fixed DAC \emph{input range} quantizes the activations and clips those beyond the range, akin
to static activation quantization in digital PTQ \cite{Nagel2021White}.

\textbf{Weights.} Weights are encoded as device conductances within a fixed range, with a
\emph{weight mapping scale} relating numerical weights to conductance values. A few large weights confine
the rest to a small part of that range. Weights are commonly represented either by the difference between two device conductances (differential mapping) or relative to an offset conductance (common-offset mapping), whose contribution is subtracted from the output \cite{Joshi2020, Xiao2022SONOS}. Both mappings are affected by programming error, read noise, and drift \cite{Joshi2020, He2021Relaxation, Xiao2022SONOS, Nandakumar2019, Mackin2022}.

With offset-based mappings, the reference contribution is computed from a nominal offset or measured using
a reference column and is subtracted before the ADC. Drift mismatch in the former, or noise in the latter, can introduce an error shared across
output columns.

\textbf{Outputs.} Digital implementations typically retain partial sums in wide
accumulators. ADC resolution is constrained by converter area and energy costs
\cite{Nagel2021White, Shafiee2016ISAAC, Murmann2024ADC}. With digital accumulation across arrays, each array's
partial output is quantized before the partial sums are combined. Thus, the resulting quantization error depends on the ADC's \emph{input range}, which sets the interval represented by its limited quantization levels. At a fixed bit width, a wider range accommodates larger outputs but increases rounding error, while a narrower range represents small outputs more precisely but suffers from clipping large outputs. All ADC bit widths in this paper are effective bits, the resolution that remains after converter noise and nonlinearity \cite{Xiao2022, Zhou2022AnalogNets}.

\textbf{Calibration.} Both the weight mapping scales and the ADC input ranges must be calibrated to achieve high accuracy in
analog CIM \cite{Joshi2020, Xiao2022, Mackin2022, Rasch2023, Lammie2024}. Many designs share one weight mapping scale and one ADC range across an array
\cite{KhaddamAljameh2022HERMES, Ambrogio2023, Yang2026PerColumn, Kneip2025IMAGINE, Zhou2022AnalogNets}, because per-column ranges need
reference circuitry comparable in area to the ADC itself, which \citet{Yang2026JSSC} call prohibitively expensive. These shared settings couple columns with
different weight and output distributions: accommodating large values reduces the resolution available to
smaller ones. Recovering accuracy while retaining this organization
avoids making column-specific calibration and control a prerequisite for deployment.

\subsection{Related work}
\label{sec:related}

Function-preserving transformations can reduce error from lossy operations (e.g., quantization) by
reparameterizing a layer without changing its ideal function. Our contribution is to construct centering
and rotations jointly for converter and device errors \emph{while retaining shared, static per-array settings},
and to account for shared components and tiled partial sums that are converted before digital accumulation
(Table~\ref{tab:error-map}).

\textbf{Rescaling.} Channel-wise scaling can trade off dynamic range between activations and weights.
SmoothQuant applies this idea to digital quantization \cite{Xiao2023SmoothQuant}. On analog arrays, it moves
activation outliers into the weights, where they set the weight mapping scale and amplify device error
(Section~\ref{sec:m-device}).

\textbf{Rotations.} Orthogonal rotations spread concentrated values across channels, lowering the peaks
that set shared quantization ranges; QuIP, QuIP\#, QuaRot, and SpinQuant use fixed or learned rotations for
low-bit digital quantization \cite{Chee2023QuIP, Tseng2024QuIPsharp, Ashkboos2024QuaRot, Liu2025SpinQuant}.
These rotations do not respect array tile boundaries, so they change every converted partial output
(Section~\ref{sec:intertile}), and their plain Hadamard transforms can concentrate common-mode errors into one
channel (Section~\ref{sec:m-adc}).

\textbf{Centering.} Subtracting fixed channel offsets and restoring their contribution via a bias can
reduce wasted range; OS+, HadaNorm, and BASE-Q combine centering with scaling and/or rotations
\cite{Wei2023OSplus, Federici2025HadaNorm, He2025BASEQ}. They target digital quantizers and inherit the
limits of the scaling or rotations they rely on.

\textbf{Analog deployment.} Prior analog methods typically target a subset of errors or relax array
constraints: NORA uses rescaling but does not address drift \cite{Hou2025NoRA}; RAELLA centers weights
but changes the array organization/arithmetic \cite{Andrulis2023RAELLA}; Ambrogio et al.\ subtract the
input mean under per-column calibration \cite{Ambrogio2023}. AdaBS recalibrates batch-normalization statistics
on data at read time to address drift error only \cite{Joshi2020}.

\section{Transforming the Analog Computation}
\label{sec:theory}
\label{sec:transforms}

Consider a linear layer $y=Wx$. For any fixed offset $\mu$ and any orthogonal matrices $R_\mathrm{out}$ and
$R_\mathrm{in}$, the layer can be rewritten exactly as
\begin{equation}
  Wx=R_\mathrm{out}^{\top}\bigl(R_\mathrm{out}WR_\mathrm{in}^{\top}\bigr)\bigl(R_\mathrm{in}x'\bigr)+W\mu,
\label{eq:reparam}
\end{equation}
where $x'=x-\mu$. On the hardware, the arrays store the transformed weights
$\widetilde W=R_\mathrm{out}WR_\mathrm{in}^{\top}$ and the DACs encode $R_\mathrm{in}x'$. The inverse rotation
$R_\mathrm{out}^{\top}$ and the precomputed term $W\mu$ are applied digitally afterward. This isometric
reparameterization computes the same function but changes the weights, inputs and partial outputs exposed to
analog error. We use it in three steps: centering removes a static input component
(Section~\ref{sec:m-static}), the output rotation reduces ADC error (Section~\ref{sec:m-adc}), and the input
rotation reduces device error (Section~\ref{sec:m-device}).

The measurements in this section use the shared
setting of Section~\ref{sec:mvm}, with one weight mapping scale and one ADC range per array.

\subsection{Static input component}
\label{sec:m-static}
\label{sec:m-centering}

The term $W\mu$ is independent of the current input. We precompute it and restore it digitally, keeping
this fixed contribution out of the noisy matrix multiplication and conversion. We choose $\mu$ as a static
per-channel mean over calibration data, which minimizes the mean squared input norm $\mathbb{E}\|x'\|_2^2$ in
the analog path without requiring dynamic calibration, and we find that it reduces ADC quantization and
clipping error.

\begin{figure}[t]
 \centering
 \includegraphics[width=\columnwidth]{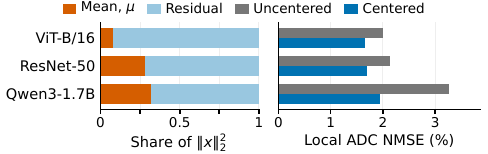}
 \caption{Static input component across all analog layers of ViT-B/16, ResNet-50 and Qwen3-1.7B,
with 4-bit ADCs and without rotations or device noise. Left: share of $\|x\|_2^2$ in the per-channel
mean. Right: local ADC error without and with centering. Bars are medians over layers.}
 \label{fig:centering-adc4}
\end{figure}

Figure~\ref{fig:centering-adc4} measures the effect on three real models. Its left panel shows the share
of $\|x\|_2^2$ in $\mu$, between a quarter and a third in ResNet-50 and Qwen3-1.7B and less in
ViT-B/16. Its right panel shows that centering lowers the local ADC error in all three.

\subsection{ADC error}
\label{sec:m-adc}
\label{sec:m-ranges}
\label{sec:m-common-mode}
\label{sec:m-structure}

With one ADC range per array, a few large outputs force a tradeoff between clipping those values and
coarsening the resolution available to the remaining outputs. An orthogonal rotation of the outputs preserves their L2 norm
but changes how it is distributed across channels.

For an untiled multiplication, omitting device and DAC error (to isolate ADC effects), folding $R_\mathrm{out}$ into the programmed
weights gives
\begin{equation}
  \widehat y=R_\mathrm{out}^{\top}\,Q_\mathrm{ADC}\bigl(R_\mathrm{out}Wx'\bigr)+W\mu,
\label{eq:pipe-out}
\end{equation}
where $Q_\mathrm{ADC}$ is the conversion.

We build $R_\mathrm{out}$ from block-diagonal Hadamard matrices. Within each block, the Hadamard coefficients have equal
magnitude, so a contribution concentrated in one channel is distributed across all transformed channels.

\begin{figure}[t]
  \centering
  \includegraphics[width=\columnwidth]{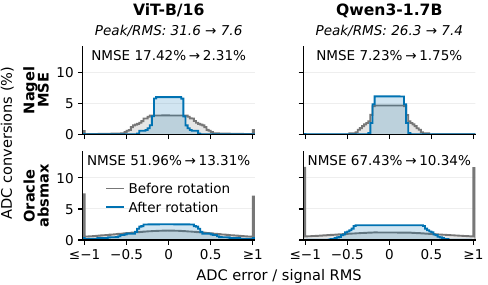}
  \caption{Output rotation reduces average local ADC error across the FFN down-projections of ViT-B/16 and Qwen3-1.7B,
with MSE-calibrated ranges (top) and ranges set from the largest output (bottom). Projections are untiled,
with 8-bit DACs, one 4-bit ADC range per projection, no centering and no device noise. NMSE is averaged and
Peak/RMS is the median over layers.}
\label{fig:adc-error}
\end{figure}

The output rotation lowers the median peak-to-RMS ratio of the ADC inputs about fourfold on the FFN
down-projections of ViT-B/16 and Qwen3-1.7B, and ADC error falls under both range-selection policies
(Figure~\ref{fig:adc-error}). The first is the MSE-calibrated ranges used throughout the paper
\cite{Nagel2021White}. The second sets ranges from the largest output, which rules out clipping but coarsens
the step.

\begin{figure}[t]
  \centering
  \includegraphics[width=\columnwidth]{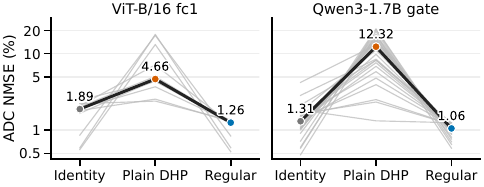}
  \caption{Regular Hadamard blocks are needed to avoid the large ADC error increase of plain blocks with random signs (DHP), which perform worse than no output rotation. Projections are untiled, with 8-bit DACs, one
MSE-calibrated 4-bit ADC range per projection and no device noise.}
\label{fig:plain-vs-regular}
\end{figure}

However, a plain
Hadamard can concentrate a shared signal before conversion, while its inverse can concentrate a shared error
afterward. Such errors arise, for example, when an imperfect reference is subtracted from every output column
in offset-based weight mappings (Section~\ref{sec:mvm}). We use regular Hadamard blocks \cite{meisner1992regular}, which satisfy
$R\mathbf{1}=R^{\top}\mathbf{1}=\mathbf{1}$ \cite{Huang2025ConvRot}, so shared components remain distributed through both the rotation
and its inverse. On the fc1 and gate projections of ViT-B/16 and Qwen3-1.7B, plain blocks with random signs (DHP) raise the median
ADC error above the unrotated computation, while regular blocks lower it (Figure~\ref{fig:plain-vs-regular}). This effect is
magnified when using fixed nominal references (Section~\ref{sec:mvm}, Appendix~\ref{app:fixedref}).

\subsection{Device error}
\label{sec:m-device}

Every device error of Section~\ref{sec:mvm}, whether from programming, reading or drift, perturbs a
conductance and, via the weight mapping scale $s$, the weight that conductance encodes. With ideal converters, the device-induced array
output error is
\begin{equation}
  e=\Delta W\,u=s\,\Delta G\,u,
\label{eq:device-error}
\end{equation}
where $u$ is the input used for that array computation.  As with ADC error, the scale depends on how far the largest weights sit above the typical
ones, so a few large weights raise $s$ for every weight that shares it. Lowering the weight peaks lowers
$s$, and with it the numerical effect of a given conductance perturbation.

The output rotation of Section~\ref{sec:m-adc} lowers the weight peaks by mixing weights across
output channels. To lower them further, we apply a fixed input Hadamard rotation $R_\mathrm{in}$, which
mixes weights across input channels and can also reduce activation peaks before the DAC.

\begin{figure}[t]
 \centering
 \includegraphics[width=\columnwidth]{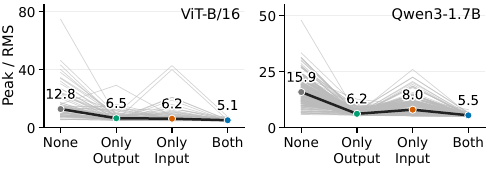}
 \caption{Weight peak-to-RMS for no rotation, output only, input only, and both, across all 72
projections of ViT-B/16 and all 196 of Qwen3-1.7B. Output and input rotations each lower weight concentration on
their own, but both are needed for the largest reduction. Faint lines are single projections, and medians are
printed.}
 \label{fig:weight-concentration}
\end{figure}

Figure~\ref{fig:weight-concentration} shows how the two rotations change weight concentration across all
projections of ViT-B/16 and Qwen3-1.7B. Each rotation alone lowers the median weight peak-to-RMS ratio,
and together they lower it by $2.5\times$ and $2.9\times$.

\begin{figure}[t]
 \centering
 \includegraphics[width=\columnwidth]{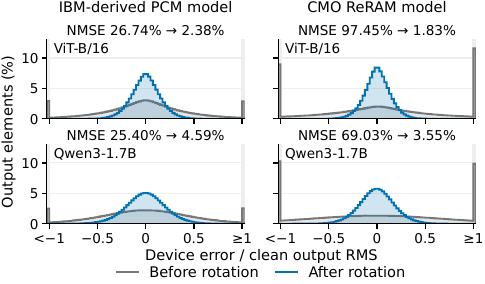}
 \caption{Local output error from device noise before and after both rotations, across the down projections of ViT-B/16 and Qwen3-1.7B, with PCM and ReRAM models at one day. Projections are untiled, with one
weight mapping scale each and no converters or centering.}
 \label{fig:device-error}
\end{figure}

Rotations also change the programmed conductances. We evaluate the combined effect of weight redistribution and conductance-dependent error
using the PCM and conductive-metal-oxide (CMO) ReRAM models of Section~\ref{sec:setup-hw}. PCM drift also introduces a shared
attenuation of the array output. Such an attenuation persists under orthogonal changes of basis, so we apply a
scalar drift-compensation gain to restore the average output scale \cite{Joshi2020}. The PCM results below
include this compensation and measure the remaining device error. Figure~\ref{fig:device-error} shows that
the two rotations substantially reduce device-only output error under both models.

\subsection{Inter-tile ADC quantization error}
\label{sec:intertile}

Layers wider than one array are split into tiles, each converted by its own ADCs before the digital sum
(Section~\ref{sec:arrays}). Input rotations crossing tile boundaries spread a
concentrated channel into every tile, widening every ADC range while accumulating ADC error. With 4-bit ADCs,
in models with concentrated down-projection inputs such as Phi-3-mini and Llama-3.2-1B, this raises the
output-normalized ADC error of the down-projections by 24\% and 62\%, and up to 12$\times$ in single layers.

We therefore keep input rotations within tiles. Each tile then computes its exact partial output $W_t x_t$, so
its ADCs quantize the same values as without input rotation and keep the same calibrated ranges.

\section{Applying the Method to Neural Networks}
\label{sec:method}
\label{sec:m-deployment}

\begin{figure*}[t]
 \centering
 \IfFileExists{figures/method/method_family_placement_combined_compact.pdf}{%
  \includegraphics[width=\textwidth]{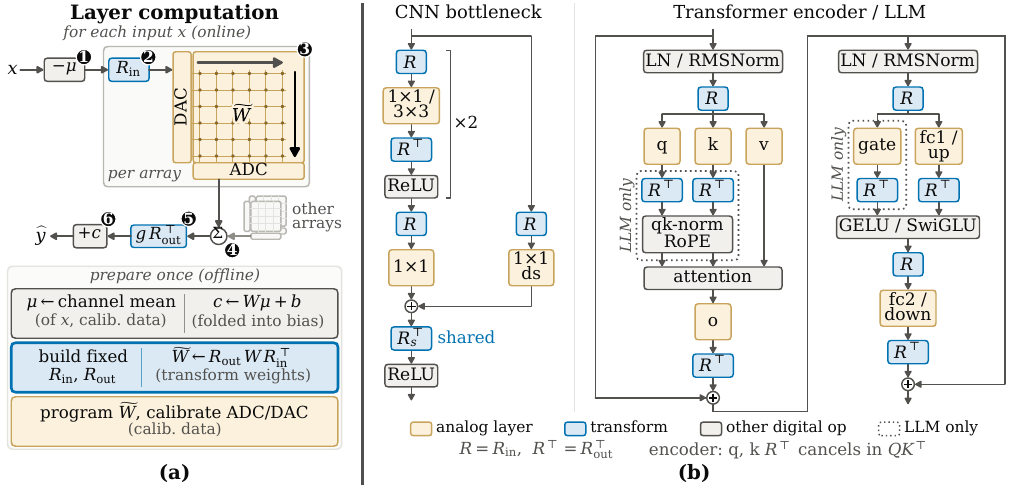}%
 }{%
  \fbox{\parbox[c][1.5in][c]{\dimexpr\textwidth-2\fboxsep-2\fboxrule\relax}{%
   \centering\small
   Figure placeholder: transformed linear layer and Hadamard placement\\[0.5em]
   CNNs \quad ViT \quad LLMs
  }}%
 }
 \caption{(a) Our method. Offline: calibrate the channel mean $\mu$, fold $c=W\mu+b$ into the bias,
build $R_\mathrm{in}$ and $R_\mathrm{out}$, transform the weights to $\widetilde W=R_\mathrm{out}WR_\mathrm{in}^\top$,
then program $\widetilde W$ and calibrate the converters. Online: (1) center, (2) rotate the inputs within each
tile, (3) compute in each array and digitize its partial output, (4) sum the digitized partial outputs, (5) undo
the output rotation and apply the drift gain $g$ (for PCM), (6) add $c$. (b) Placement of the rotations in a CNN bottleneck and a transformer block (ViT or
LLM).}
 \label{fig:method-family-placement}
\end{figure*}

\subsection{Our method on a single layer}
\label{sec:m-preparation}

Figure~\ref{fig:method-family-placement}a gives the preparation and execution of a transformed layer.

Every operation the method adds is an affine map. Centering subtracts a fixed vector from the
input, and its correction folds into the existing bias, or adds one where the layer has none. We use one compensation gain per layer and fold it
into the inverse output rotation's scaling (Appendix~\ref{app:drift-gain}).
Where possible, we use four-point add/subtract kernels, which cost $\tfrac{7}{8}\log_2 b$ additions per element
for block size $b$ \cite{coppersmith1994hadamard}. The normalization
reduces to shifts or folds into existing scales, so the transforms themselves need no multiplications.
Section~\ref{sec:r-cost} reports the measured digital cost.

\begin{table*}[t]
\centering
\footnotesize
\setlength{\tabcolsep}{3pt}
\begin{minipage}[t]{0.50\textwidth}
\centering
\caption{Evaluated models and tasks. All LLM runs keep the opening template token digital
(Appendix~\ref{app:sink}).}
\label{tab:setup}
\begin{tabular}{@{}p{0.11\linewidth}p{0.50\linewidth}p{0.33\linewidth}@{}}
\toprule
Family & Models & Task and metric \\
\midrule
CNNs & ResNet-18, ResNet-50 \cite{He2016ResNet} & ImageNet top-1 \cite{Russakovsky2015ImageNet} \\
ViTs & ViT-B/32, ViT-B/16, ViT-L/16, DeiT-S \cite{Dosovitskiy2021ViT, Touvron2021DeiT} & ImageNet top-1 \\
LLMs & Llama-3.2-1B, Qwen3-1.7B, Qwen3-8B, Phi-3-mini, Phi-4 \cite{Grattafiori2024Llama3, Yang2025Qwen3,
  Abdin2024Phi3, Abdin2024Phi4} & Mean of MMLU, HellaSwag, ARC-E and ARC-C \cite{Hendrycks2021MMLU,
  Zellers2019HellaSwag, Clark2018ARC} \\
\bottomrule
\end{tabular}
\end{minipage}\hfill
\begin{minipage}[t]{0.46\textwidth}
\centering
\caption{Default hardware setting.}
\label{tab:setup-hw}
\begin{tabular}{@{}p{0.24\linewidth}p{0.72\linewidth}@{}}
\toprule
Setting & Value \\
\midrule
Arrays & $512\times512$, one weight mapping scale and one ADC range each \\
Converters & 4-, 6- or 8-bit ADCs, and a static 8-bit DAC with one range per tensor \cite{Buchel2025AFM} \\
Devices & PCM and CMO ReRAM with one reference column per array, read one day after programming
  \cite{Rasch2021aihwkit, Nandakumar2019, Falcone2025} \\
Weights & Continuous conductances, with an effective precision of about 4 bits at one day
  (Appendix~\ref{app:eff-wbits}) \\
Drift compensation & Global, on PCM: one gain per layer for our method, and one per array for the baselines,
  following \citet{Joshi2020}'s online read procedure (Appendix~\ref{app:drift-gain}).\\
\bottomrule
\end{tabular}
\end{minipage}
\end{table*}

\subsection{Application to model families}
\label{sec:m-families}

\label{sec:m-placement}

The input and output layers of every model remain digital \cite{Joshi2020, Rasch2023, Buchel2025AFM}.

\textbf{CNNs.} Rotation-based quantization has targeted transformers \cite{Ashkboos2024QuaRot,
Liu2025SpinQuant}. We also apply the transforms to CNNs: we fold BatchNorm into convolution weights \cite{Nagel2021White} and apply the method to residual-block
convolutions and projection shortcuts. Sharing the input channel transform across kernel taps
allows reuse across overlapping patches. At projection shortcuts, both branches use the same
output transform, allowing a single inverse after their sum and before the activation.

\textbf{Transformer models.} These placements follow the computational-invariance recipe of SliceGPT and
QuaRot \cite{Ashkboos2024SliceGPT, Ashkboos2024QuaRot, Liu2025SpinQuant}. Attention and normalization
remain digital. V remains transformed
through attention, with its inverse folded into the output projection's weights. The first
feed-forward projections are restored before their nonlinearity and gating, and the attention-output
and second feed-forward projections are inverted before the residual addition.

\textbf{ViTs.} Q and K use matching head-local transforms, preserving their
attention scores in exact arithmetic without an explicit inverse \cite{Ashkboos2024QuaRot}.

\textbf{LLMs.} Q and K are restored after their projection ADCs, before any Q/K
normalization and rotary positional encoding. Grouped-query attention repeats each KV head's transform
across its associated query heads.

Like PrefixQuant and CushionCache \cite{Chen2024PrefixQuant, Son2024CushionCache}, we preload
the key-value cache with the attention-sink token \cite{xiao2024streamingllm, sun2024massive} that opens every prompt, with no additional per-token arithmetic or hardware change.
This keeps its outliers out of the static DAC ranges (Appendix~\ref{app:sink}).

\section{Evaluation setup}
\label{sec:setup}
\label{sec:setup-hw}

\begin{figure*}[t]
\centering
\includegraphics[width=\textwidth]{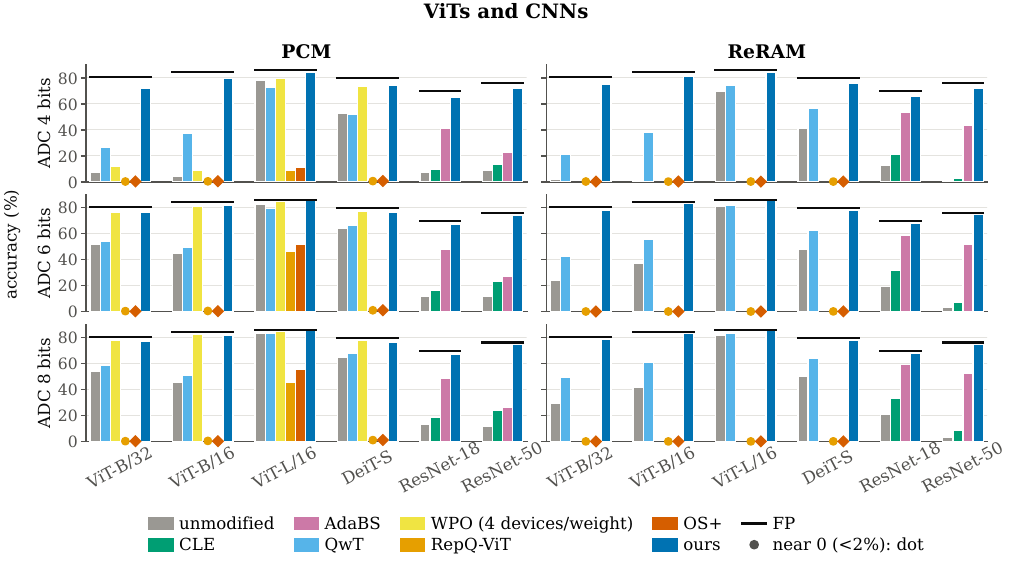}
\caption{ImageNet top-1 accuracy of ViTs and CNNs on PCM and ReRAM with N-bit ADCs, 1 day
after programming.}
\label{fig:s51-encoders}
\end{figure*}

Table~\ref{tab:setup} lists the evaluated models and tasks, and Table~\ref{tab:setup-hw} the hardware
setting that Section~\ref{sec:results} uses unless stated otherwise. All other simulation parameters use the
AIHWKit defaults \cite{Rasch2021aihwkit}. LLM accuracy is the mean over the four benchmarks,
evaluated with lm-eval \cite{Gao2024LMEval}, and calibration and evaluation data are disjoint for every model.
All methods share the calibration data and the ADC range rule, an MSE fit \cite{Nagel2021White} on clean
statistics, scaled once before deployment by the device model's nominal drift at a target read age, without
on-chip measurement. The target equals the read age unless stated otherwise. Static DAC ranges come from a clipping histogram
observer. Transforms are fixed per seed
(Appendix~\ref{app:eval}).

\section{Results}
\label{sec:results}

\subsection{ViTs and CNNs}
\label{sec:r-main}

\subsubsection{Headline results}

Figure~\ref{fig:s51-encoders} shows ViT and CNN results against post-training baselines on PCM and ReRAM with
4- to 8-bit ADCs. The unmodified models collapse in many cases, due to device and ADC error. Our method is more
robust. With 6- and 8-bit ADCs, it reaches within 4.2 percentage points of floating-point accuracy for every
model. Furthermore, it even works with 4-bit ADCs, further reducing converter area and energy, at an average
additional cost of 2.3 percentage points.

The baselines fall into four loose groups.
\begin{itemize}
\item Digital PTQ methods such as OS+ \cite{Wei2023OSplus} and RepQ-ViT \cite{Li2023RepQViT} shift magnitude
  from activations to weights within a channel, which worsens the weight mapping scale mismatch. They collapse in
  most settings.
\item CLE \cite{Nagel2019DFQ} and AdaBS \cite{Joshi2020} recover CNN accuracy but cannot fully close the gap.
\item QwT \cite{Fu2025QwT} is an adapted digital method that uses a large number of digital operations to recover
  accuracy, yet trails our method.
\item WPO \cite{Mackin2022} performs well against device noise, but cannot correct ADC error and fails with 4-bit
  ADCs on ViT-B/16 and ViT-B/32. It also requires four devices per weight, a significant penalty, while our
  method uses simple shared arrays.
\end{itemize}

\subsubsection{Comparison with hardware-aware training}
\label{sec:r-main-hwa}

We compare ResNet-18 and ResNet-50 with the published HWA
results of \citet{Rasch2023} in our reproduction of their setting (Appendix~\ref{app:hwa-setup}). Despite no retraining, our method
comes very close to their results on PCM and matches them on ReRAM (Table~\ref{tab:s57-hwa}).

\begin{table}[t]
\centering
\footnotesize
\setlength{\tabcolsep}{3pt}
\caption{ResNet accuracy (\%) alongside published HWA results \cite{Rasch2023}, with PCM or the ReRAM model
of \citet{Wan2022}, 8-bit ADCs and one-day read age, with per-column weight mapping scales as in \citet{Rasch2023}. Unmodified and Ours report mean $\pm$ SD over three seeds.}
\label{tab:s57-hwa}
\begin{tabular}{@{}llrrr@{}}
\toprule
Model & Device & \textit{Unmodified} & HWA & \textbf{Ours} \\
\midrule
ResNet-50 & PCM   & $\mathit{61.3\pm3.1}$ & 74.7 & $\mathbf{73.7\pm0.1}$ \\
ResNet-18 & PCM   & $\mathit{56.1\pm0.8}$ & 68.0 & $\mathbf{67.5\pm0.1}$ \\
ResNet-50 & ReRAM & $\mathit{58.1\pm1.9}$ & 72.4 & $\mathbf{72.8\pm0.1}$ \\
ResNet-18 & ReRAM & $\mathit{44.9\pm3.3}$ & 65.1 & $\mathbf{65.6\pm0.2}$ \\
\bottomrule
\end{tabular}
\end{table}

\subsection{LLMs}
\label{sec:r-llm}
\label{sec:r-hwa}

\begin{figure}[t]
\centering
\includegraphics[width=\columnwidth]{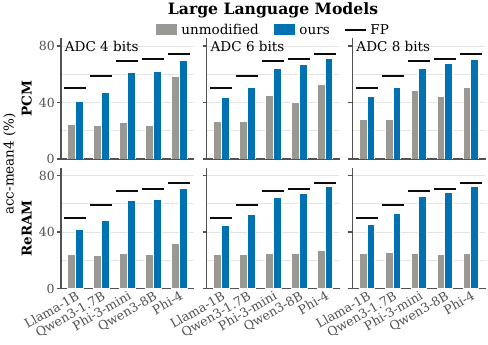}
\caption{LLM mean accuracy over MMLU, HellaSwag, ARC-Easy and ARC-Challenge on PCM and ReRAM with 4-, 6-
and 8-bit ADCs, one day after programming. Lines mark floating-point accuracy.}
\label{fig:s51-decoders}
\end{figure}

\subsubsection{Headline results}

Figure~\ref{fig:s51-decoders} shows LLM results on PCM and ReRAM with 4- to 8-bit ADCs. The unmodified models
fall to near-chance accuracy in most settings. Our method is effective at improving accuracy, with an average
gain of 28.1 percentage points, but cannot fully recover it for small LLMs. It becomes more effective with
scale: Qwen3-8B and Phi-4 (14B) come within 4 percentage points of floating-point accuracy with 6- and 8-bit
ADCs, while the smaller Llama-3.2-1B and Qwen3-1.7B keep larger gaps. With 4-bit ADCs, accuracy only drops by
another 3.0 percentage points on average. QuaRot \cite{Ashkboos2024QuaRot}, applied to the arrays directly,
trails our method by up to 44.3 percentage points
(Appendix~\ref{app:quarot}).

\subsubsection{Comparison with hardware-aware training}
\label{sec:r-llm-hwa}

We compare Phi-3-mini and Llama-3.2-1B with the analog foundation models (AFM) of \citet{Buchel2025AFM} in
our reproduction of their setting (Appendix~\ref{app:hwa-setup}). Despite no retraining, our method comes very close to AFM on
Phi-3-mini and within 2.7 percentage points on Llama-3.2-1B (Table~\ref{tab:s57-afm}). Per-task results and the
comparison with the baselines of \citet{Buchel2025AFM} are in Appendix~\ref{app:hwa-setup}.

\begin{table}[t]
\centering
\footnotesize
\setlength{\tabcolsep}{3pt}
\caption{LLM mean accuracy (\%) compared with the published AFM results \cite{Buchel2025AFM}, in our
reproduction of their setting, with their checkpoints, scoring and noise level (Appendix~\ref{app:hwa-setup}).
Unmodified and Ours report mean $\pm$ SD over ten seeds.}
\label{tab:s57-afm}
\begin{tabular}{@{}lr|r|r@{}}
\toprule
Model & \textit{Unmodified} & AFM & \textbf{Ours} \\
\midrule
Phi-3-mini   & $\mathit{72.9\pm0.9}$ & 79.7 & $\mathbf{78.4\pm0.5}$ \\
Llama-3.2-1B & $\mathit{36.4\pm1.6}$ & 45.5 & $\mathbf{42.8\pm1.5}$ \\
\bottomrule
\end{tabular}
\end{table}

\subsection{Component contributions}
\label{sec:r-ablation}

\begin{figure}[t]
\centering
\includegraphics[width=\columnwidth]{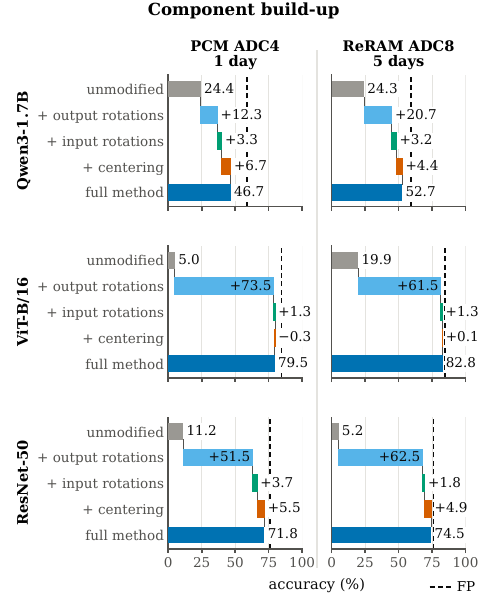}
\caption{Accuracy as output rotations, input rotations and centering are added in that order, on PCM with
4-bit ADCs at one day and ReRAM with 8-bit ADCs at five days. Dashed lines mark floating-point accuracy.}
\label{fig:s52-waterfall}
\end{figure}

We ablate the method on ViT-B/16, ResNet-50 and Qwen3-1.7B in two settings, one dominated by ADC error and one
by device error: PCM with 4-bit ADCs one day after programming, and ReRAM with 8-bit ADCs five days after
programming. Appendix~\ref{app:loo} removes one component at a time.

We find that output rotations provide the bulk of the gain in every setting, especially for ViT-B/16
(Figure~\ref{fig:s52-waterfall}). As Section~\ref{sec:theory} discusses, they reduce both ADC and device
error. However, input rotations and centering are needed to approach floating-point accuracy, especially for
ResNet-50 and Qwen3-1.7B. This suggests that for some models and operating conditions, the method does not have
to be applied in full to obtain most of its benefit, saving digital operations.

\subsection{Overheads}
\label{sec:r-cost}
\label{sec:r-ladder}
\label{sec:r-age}

\begin{figure}[t]
\centering
\includegraphics[width=\columnwidth]{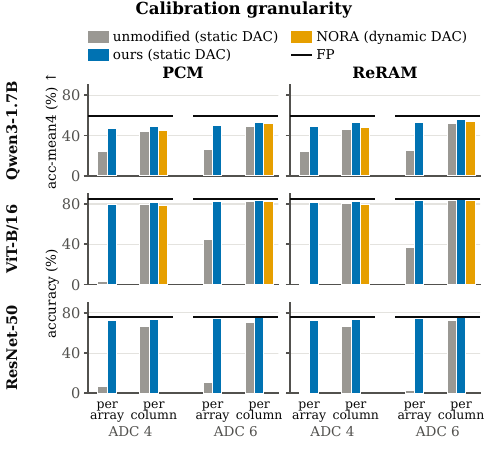}
\caption{Accuracy with per-array or per-column weight mapping scales and ADC ranges, one day after
programming. NORA uses its native dynamic DAC.}
\label{fig:s53-granularity}
\end{figure}

\begin{figure}[t]
\centering
\includegraphics[width=\columnwidth]{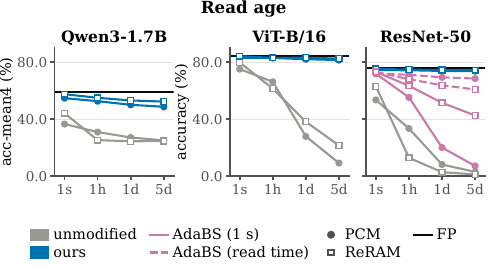}
\caption{Accuracy against read age with 6-bit ADCs. The arrays are programmed and calibrated once at 1\,s
and not refreshed.}
\label{fig:s56-read-age}
\end{figure}

\begin{figure}[t]
\centering
\includegraphics[width=\columnwidth]{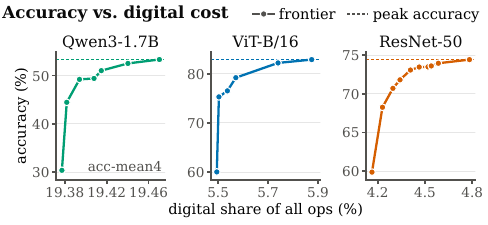}
\caption{Accuracy against the digital share of all operations on ReRAM with 8-bit ADCs at five days. Each curve
is the frontier of the evaluated configurations, and dotted lines mark its peak accuracy. The unmodified models lie below the plotted range.}
\label{fig:s52-frontier-total}
\end{figure}

On analog accelerators, digital operations run on separate compute cores \cite{Jain2023Mesh} and can become the
energy and area bottleneck \cite{Spoon2021}. Our method adds few of them and also lowers analog cost in two ways:
the arrays can be simpler, and even simple arrays need to be refreshed less often.

\textbf{Calibration granularity.} Per-column weight mapping scales and ADC ranges need peripheral circuitry
that can be prohibitively expensive (Section~\ref{sec:mvm}). With one scale and one range per array, our method matches or exceeds
the unmodified model with per-column settings, by up to 6.0 percentage points (Figure~\ref{fig:s53-granularity}). With per-column settings, it adds up to 7.1 points. For fairness, we also
include NORA \cite{Hou2025NoRA}, an analog-targeted baseline that needs per-column settings and a dynamic DAC.
With one setting per array and a static DAC, our method stays within 1.6 points of NORA, and it exceeds
NORA with per-column settings.

\textbf{Refresh.} Drift accumulates after programming, and refreshing the arrays to reset it costs hardware
and system resources. We program and calibrate the arrays once at 1\,s and read them up to five days later
without refresh. Our method loses at most 1.7 percentage points on the vision models over this period, while
the unmodified models collapse (Figure~\ref{fig:s56-read-age}). Qwen3-1.7B loses more with age but keeps a
large advantage. AdaBS \cite{Joshi2020}, calibrated once at 1\,s, falls to 7.2\% (PCM) and 42.7\% (ReRAM)
on ResNet-50 after five days. Even when recalibrated at read time, it stays 5.2 and 13.5 points below our method.
Arrays running our method can therefore be refreshed less often.

\textbf{Digital cost.} We trade accuracy for digital operations by reducing the
Hadamard block size and by dropping input or output rotations from some layer types. Appendix~\ref{app:qwt-cost} gives the configurations and the
counting method. The evaluated configurations form a frontier (Figure~\ref{fig:s52-frontier-total}). Accuracy rises steeply with the first transforms and
then saturates, and even the most accurate configurations add less than 1\% additional operations during
model execution. Together, these figures span a design space of cost and accuracy.

\subsection{DAC precision}
\label{sec:r-dac}
\label{app:converter-grid}

\begin{figure}[t]
\centering
\includegraphics[width=\columnwidth]{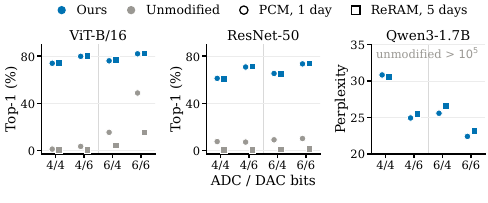}
\caption{Accuracy with 4- and 6-bit ADCs and DACs on PCM at one day and ReRAM at five days: top-1 accuracy
for ViT-B/16 and ResNet-50, and WikiText-2 \cite{merity2017wikitext} perplexity for Qwen3-1.7B (lower is better). ViT-B/16 floating-point accuracy is 84.6\%, ResNet-50 76.1\%.}
\label{fig:converter-grid}
\end{figure}

The main results use 8-bit static DACs. As a case study, we also vary DAC and ADC precision between 4 and 6 bits (Figure~\ref{fig:converter-grid}). With 4-bit DACs and
4-bit ADCs, our method keeps 74.1--74.3\% on ViT-B/16, 61.0--61.2\% on ResNet-50 and a perplexity of about 31
on Qwen3-1.7B, while the unmodified models collapse.
As in digital post-training quantization, the input rotation spreads activation outliers and lowers the
range each DAC must cover (Section~\ref{sec:m-device}).

\subsection{Partially weight-stationary SRAM CIM}
\label{sec:r-sram}

\begin{figure}[t]
\centering
\includegraphics[width=\columnwidth]{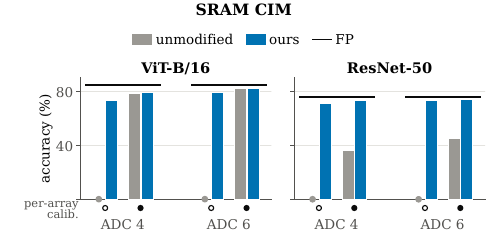}
\caption{SRAM CIM accuracy with one weight mapping scale and one ADC range per tensor (open dots, as in PWS) or
per array (filled dots, as in FWS). Dots at zero indicate accuracies below 2\%.}
\label{fig:s55-pws}
\end{figure}

PWS SRAM CIM arrays load many weight tiles in turn (Section~\ref{sec:arrays}), so they are typically calibrated
once per tensor rather than per tile. Inspired by PICO-RAM \cite{Chen2025PICORAM}, we simulate signed 4-bit
weights in arrays of 144 rows and 8 outputs, 8-bit DACs, and 4- or 6-bit ADCs, with weight and converter
quantization only (Figure~\ref{fig:s55-pws}). With one setting per tensor, the unmodified models collapse, while
our method reaches 71.3--79.4\%. On ResNet-50, it even exceeds the unmodified model with the per-array settings
of FWS arrays, by 28 to 35 points.

\section{Conclusion}
\label{sec:conclusion}

We showed that the analog error of a layer depends on the representation in which it is computed, and that
this representation can be chosen offline through matched isometric reparameterizations. Centering, regular
output rotations and tile-local input rotations reduce converter and device error on unmodified PCM and ReRAM
arrays with one weight mapping scale and one ADC range each. With 6- and 8-bit ADCs, CNNs and ViTs come within
4.2 percentage points of floating-point accuracy, and LLMs gain substantially, more so as they grow. Compared
with published hardware-aware training, the approach matches it or comes within 2.7 points. It adds less than
1\% to the operations of model execution, provides the accuracy that per-column calibration would otherwise
require, and allows less frequent refresh. The focus of this work has been to show that analog CIM without
retraining is achievable, as \citet{Buchel2025AFM} called for. Co-optimizing the array design with these
transforms, quantifying their system-level impact, and closing the remaining gap for small LLMs are left for
future work.

\clearpage
\bibliography{refs}
\bibliographystyle{mlsys2025}

\clearpage
\AtAppendixStart
\appendix
\twocolumn
\raggedbottom
\setlength{\intextsep}{7pt}
\setlength{\textfloatsep}{10pt}
\renewcommand{\topfraction}{0.95}
\renewcommand{\bottomfraction}{0.85}
\renewcommand{\textfraction}{0.05}

\section{Evaluation details}
\label{app:eval}\label{sec:config}

\textbf{Models and data.} The ViTs use the timm AugReg checkpoints, pretrained on ImageNet-21k and
fine-tuned on ImageNet-1k \cite{Steiner2022AugReg}. DeiT-S and the ResNets use their ImageNet-1k checkpoints
from timm and torchvision. For the vision models, we use 1,024 ImageNet validation images for calibration and
the remaining 48,976 for evaluation. Ablations and converter studies use 4,898 evaluation images. CNN and ViT
centering means use 1,024 separate training images. LLM accuracy is the unweighted mean of MMLU, HellaSwag,
ARC-Easy, and ARC-Challenge, using likelihood scoring. Calibration uses approximately 8,192 held-out tokens per
task. The opening chat-template token is digital and unscored.

\textbf{Calibration and refresh.} All static settings come from a clean calibration pass before deployment:
a simulated forward pass over the calibration data without any analog effects. The ADC ranges are then scaled
once, offline, by the device model's nominal drift at the target read age. At read time, only the PCM drift gain
is measured, following \citet{Joshi2020} (Table~\ref{tab:app-calib}). Because the rotations make this gain nearly
uniform across the arrays of a layer (Appendix~\ref{app:drift-gain}) and make the computation on the arrays
itself more robust to drift (Section~\ref{sec:r-age}), we expect them to assist drift compensation
without online reads. We did not test this and use the default AIHWKit implementation \cite{Rasch2021aihwkit}.

\begin{table}[H]
\centering\footnotesize\setlength{\tabcolsep}{4pt}
\caption{What is calibrated, measured or refreshed.}
\label{tab:app-calib}
\begin{tabular}{@{}lp{0.68\columnwidth}@{}}
\toprule
Input mean $\mu$ & Clean pass, per input channel \\
DAC range & Clean pass, per layer input \\
ADC range & Clean pass, per array, scaled offline for nominal drift \\
Drift gain $g(t)$ & PCM only, probe readout at read time \cite{Joshi2020} \\
Reference column & ReRAM only, one per array, subtracted before the ADC \\
Weights & Programmed once at $t=0$, never refreshed \\
\bottomrule
\end{tabular}
\end{table}

\textbf{Weight mapping and converters.} Each array's largest absolute weight maps to maximum
conductance. Main noisy-device runs use continuous targets. Static DAC ranges use a histogram observer, and
ADCs use signed, symmetric quantization with MSE-based range selection. LLM down-projection DAC ranges are
clipped at the 99.9th percentile.

\textbf{Transforms.} Output rotations use regular Hadamard blocks of the largest power of four up to 256
that divides the layer width, and 64 in the ResNets and the 384-wide DeiT-S layers. Input rotations use plain
Hadamard blocks within each tile. Attention projections use head-local plain blocks, including three blocks of
32 per head in Phi-3-mini. All transforms are built in closed form with fixed signs and permutations.

\textbf{Simulator.} Most evaluation scripts build on AIHWKit \cite{Rasch2021aihwkit} at commit eb089df.

\section{Digital cost}\label{app:qwt-cost}

\textbf{Added operations.} Each Hadamard transform of block size $b$ costs $\tfrac{7}{8}\log_2 b$ additions per
element with the four-point construction, which we use whenever possible, and at most $\log_2 b$ otherwise, and no
multiplications (Section~\ref{sec:m-preparation}). Centering adds one subtraction per input
element of each analog projection. The centering correction and the drift gain fold into existing digital
operations and add no cost, except in Qwen3-1.7B, which has no biases and adds one addition per output for the centering correction.

\textbf{Normalization.} We report the added operations as a share of all operations, where each analog
multiply-accumulate counts as two operations and the existing digital operations cover the whole model,
including its digital input and output layers. Costs are per token for ViT (image cost divided by 197) and
Qwen3-1.7B (at a reference context of 389.5 tokens), and per image for ResNet.
Figures~\ref{fig:s52-frontier-total} and~\ref{fig:s52-noise} plot the total digital share $(B+E)/(A+B+E)$,
where $A$ counts each analog multiply-accumulate as two operations, $B$ is the existing digital work and $E$ the
added operations. The added share alone is $E/(A+B+E)$.
Existing digital operations comprise the attention products ($QK^\top$, $AV$), softmax with score
scaling, LayerNorm and RMSNorm (including the Q/K head norms), activations (GELU, SwiGLU), RoPE, residual
additions, DAC quantizers, accumulation across arrays, and the digital input and output layers. Each scalar
operation counts as one. Signs, permutations, shifts and data movement are free.

\textbf{Frontier configurations.} The frontier evaluates partial applications of the method:
rotations restricted to subsets of layer types, with and without centering, at block sizes from 4 to 256.
Each point is costed from its own configuration.

\medskip

\section{Precomputed first token}\label{app:sink}
\begin{figure}[H]
\centering
\includegraphics[width=\linewidth]{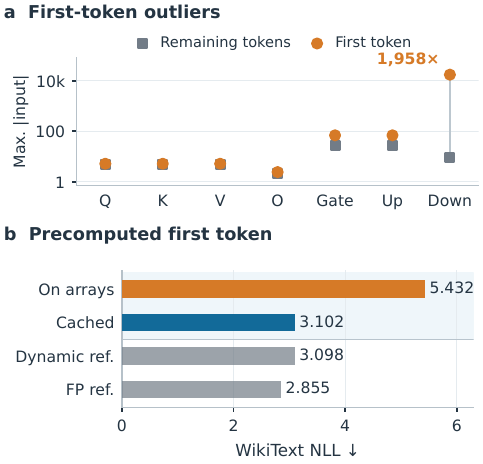}
\caption{Qwen3-1.7B first-token diagnostic. (a) Maximum absolute layer-2 projection inputs for
the first and remaining tokens over four prompts. (b) WikiText NLL, the log perplexity, on PCM with 6-bit ADCs,
8-bit DACs and a one-day read age, showing the impact of executing the first token on the arrays.}
\label{fig:app-sink}
\end{figure}

\textbf{First-token outlier.} Prompts share an opening chat-template token that receives
63.3\% of attention in the clean diagnostic. Its layer-2 down-projection input reaches
17,616 versus 9.0 for remaining tokens ($1{,}958\times$, Figure~\ref{fig:app-sink}a).
Including it in the histogram observer widens the static DAC range by $3.83\times$.

\textbf{Precomputation.} We remove the clean opening token from array execution and range
calibration, caching its hidden state and K/V entries. The prefix remains in subsequent attention. This does not require any online operations, as it can be precomputed for causal attention.

\textbf{Result.} Precomputation lowers NLL from 5.4319 to 3.1018 in the matched static-DAC
comparison (Figure~\ref{fig:app-sink}b). Calibration uses four consecutive 2,048-token
WikiText-2 \cite{merity2017wikitext} training windows, and evaluation uses 32 test windows, excluding the prefix and first
content token from scoring.
\begin{table*}[t]
\centering
\caption{AFM comparison: per-task accuracy (\%) and the unweighted four-task mean.
$k$ multiplies the programming-noise standard deviation: 1.00 is nominal and 1.10
is adjusted to match the published noise-only baseline.
Our SI8/O8 rows use one seed at $k=1.00$ and ten at $k=1.10$, and noise-only rows use
four Phi seeds and three/ten Llama seeds at $k=1.00/1.10$.
Uncertainties are sample SDs. SI/DI denote static/dynamic inputs, and
W16h/W4h denote the publication's hardware-noisy weight formats.}
\label{tab:afm-comp}
\label{tab:app-afm-nominal}
\begingroup
\fontsize{8.5}{10.5}\selectfont
\setlength{\tabcolsep}{2.5pt}
\renewcommand{\arraystretch}{1.12}
\begin{tabular*}{\textwidth}{@{\extracolsep{\fill}}llcrrrrr@{}}
\toprule
Method & Precision & $k$ & MMLU & HellaSwag & ARC-C & ARC-E & Mean \\
\midrule
\multicolumn{8}{@{}l}{\textbf{Phi-3-mini}} \\
\multicolumn{8}{@{}l}{Published results} \\
FP & W16 & -- & 69.4 & 83.5 & 84.6 & 91.1 & 82.1 \\
Original & W16h & -- & 63.3 & 73.4 & 80.4 & 89.0 & 76.5 \\
AFM & SI8-W16h-O8 & -- & 65.1 & 80.6 & 83.2 & 89.8 & 79.7 \\
LLM-QAT \cite{liu2024llmqat} & SI8-W4h & -- & 60.1 & 74.8 & 77.9 & 87.2 & 75.0 \\
SpinQuant & SI8-W4h & -- & 29.7 & 25.1 & 29.6 & 36.0 & 30.1 \\
SpinQuant & DI8-W4h & -- & 48.3 & 33.4 & 60.6 & 75.7 & 54.5 \\
\cmidrule(lr){1-8}
\multicolumn{8}{@{}l}{\textit{Our measurements}} \\
\textit{FP} & \textit{W16} & \textit{--} & \textit{68.9} & \textit{83.8} & \textit{84.9} & \textit{91.5} & \textit{82.3} \\
\textit{Noise only} & \textit{W16h} & \textit{1.00} & \textit{64.3} & \textit{76.0} & \textit{81.7} & \textit{89.5} & \textit{77.9} \\
\textit{Noise only} & \textit{W16h} & \textit{1.10} & \textit{63.1\,\(\pm\)\,0.4} & \textit{73.4\,\(\pm\)\,2.3} & \textit{80.7\,\(\pm\)\,0.5} & \textit{88.8\,\(\pm\)\,0.6} & \textit{76.5\,\(\pm\)\,0.9} \\
\textit{Unmodified} & \textit{SI8-W16h-O8} & \textit{1.00} & \textit{60.4} & \textit{72.5} & \textit{81.0} & \textit{88.6} & \textit{75.6} \\
\textit{Unmodified} & \textit{SI8-W16h-O8} & \textit{1.10} & \textit{58.5\,\(\pm\)\,0.5} & \textit{66.8\,\(\pm\)\,2.8} & \textit{78.6\,\(\pm\)\,1.0} & \textit{87.7\,\(\pm\)\,0.7} & \textit{72.9\,\(\pm\)\,0.9} \\
\textbf{\textit{Ours}} & \textit{SI8-W16h-O8} & \textit{1.00} & \textit{64.9} & \textit{79.6} & \textit{82.3} & \textit{90.4} & \textit{79.3} \\
\textbf{\textit{Ours}} & \textit{SI8-W16h-O8} & \textit{1.10} & \textit{63.7\,\(\pm\)\,0.5} & \textit{77.3\,\(\pm\)\,1.9} & \textit{82.4\,\(\pm\)\,0.6} & \textit{90.2\,\(\pm\)\,0.4} & \textit{78.4\,\(\pm\)\,0.5} \\
\addlinespace[3pt]
\midrule
\multicolumn{8}{@{}l}{\textbf{Llama-3.2-1B}} \\
\multicolumn{8}{@{}l}{Published results} \\
FP & W16 & -- & 46.9 & 33.2 & 51.7 & 69.2 & 50.2 \\
Original & W16h & -- & 35.9 & 26.5 & 39.2 & 52.4 & 38.5 \\
AFM & SI8-W16h-O8 & -- & 42.9 & 29.3 & 46.7 & 63.1 & 45.5 \\
LLM-QAT & SI8-W4h & -- & 40.1 & 29.0 & 43.8 & 59.0 & 42.9 \\
SpinQuant & SI8-W4h & -- & 25.6 & 24.9 & 24.8 & 24.7 & 25.0 \\
SpinQuant & DI8-W4h & -- & 33.2 & 25.8 & 35.4 & 46.1 & 35.1 \\
\cmidrule(lr){1-8}
\multicolumn{8}{@{}l}{\textit{Our measurements}} \\
\textit{FP} & \textit{W16} & \textit{--} & \textit{46.0} & \textit{32.5} & \textit{50.6} & \textit{68.6} & \textit{49.4} \\
\textit{Noise only} & \textit{W16h} & \textit{1.00} & \textit{38.4} & \textit{27.0} & \textit{40.0} & \textit{57.2} & \textit{40.7} \\
\textit{Noise only} & \textit{W16h} & \textit{1.10} & \textit{36.6\,\(\pm\)\,1.4} & \textit{26.3\,\(\pm\)\,0.8} & \textit{37.9\,\(\pm\)\,1.8} & \textit{52.2\,\(\pm\)\,4.0} & \textit{38.2\,\(\pm\)\,1.5} \\
\textit{Unmodified} & \textit{SI8-W16h-O8} & \textit{1.00} & \textit{37.4} & \textit{27.3} & \textit{38.9} & \textit{53.7} & \textit{39.3} \\
\textit{Unmodified} & \textit{SI8-W16h-O8} & \textit{1.10} & \textit{35.1\,\(\pm\)\,1.4} & \textit{26.0\,\(\pm\)\,0.7} & \textit{36.1\,\(\pm\)\,2.5} & \textit{48.3\,\(\pm\)\,3.7} & \textit{36.4\,\(\pm\)\,1.6} \\
\textbf{\textit{Ours}} & \textit{SI8-W16h-O8} & \textit{1.00} & \textit{41.6} & \textit{27.6} & \textit{46.2} & \textit{62.5} & \textit{44.5} \\
\textbf{\textit{Ours}} & \textit{SI8-W16h-O8} & \textit{1.10} & \textit{40.3\,\(\pm\)\,1.3} & \textit{28.2\,\(\pm\)\,1.1} & \textit{43.8\,\(\pm\)\,2.8} & \textit{58.9\,\(\pm\)\,2.7} & \textit{42.8\,\(\pm\)\,1.5} \\
\bottomrule
\end{tabular*}
\endgroup
\end{table*}

\section{Leave-one-out ablation}
\label{app:loo}

\begin{figure}[H]
\centering
\includegraphics[width=\columnwidth]{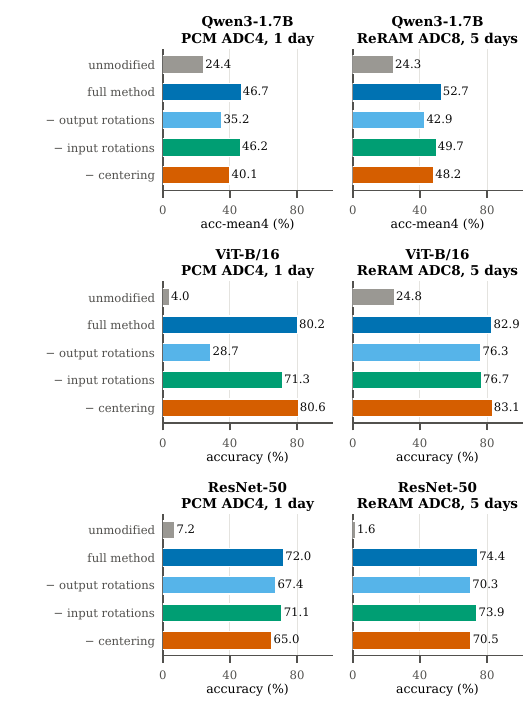}
\caption{Accuracy with one component removed from our method, on PCM with 4-bit ADCs at one day and ReRAM with
8-bit ADCs at five days. Vision models use 4,898 evaluation images and one seed.}
\label{fig:app-loo}
\end{figure}

Removing one component at a time from the full method confirms the build-up of
Section~\ref{sec:r-ablation} (Figure~\ref{fig:app-loo}). Output rotations matter most, except for ResNet-50 on PCM, where centering matters more. Of the other
two components, input rotations matter more for ViT-B/16, and centering for ResNet-50 and Qwen3-1.7B. For ViT-B/16, removing centering
changes accuracy by less than 0.5 points.

\section{Centering on ViT-B/16}\label{app:vit-centering}

Centering barely changes ViT-B/16 accuracy once both rotations are applied, by $-0.5$ to $+0.1$ points across
devices and ADC precisions. Its inputs carry little static component, a median of 8\% of their squared L2 norm against
28\% in ResNet-50 and 32\% in Qwen3-1.7B (Section~\ref{sec:m-static}). The output rotation also spreads the
static output that remains across columns, so centering leaves little to remove: the median per-projection ratio of uncentered to centered 4-bit
ADC error is only $1.06$, against $1.35$ in ResNet-50 and $1.24$ in Qwen3-1.7B.

\section{Plain and regular blocks with a fixed reference}
\label{app:fixedref}

CMO ReRAM arrays can subtract a fixed nominal reference instead of a reference column, as AIHWKit does by
default \cite{Rasch2021aihwkit}, which saves the reference column and its tracking circuitry. The relaxation
shift is then not cancelled and adds a shared error to every output column. We do not use this default for our
results, because the unmodified model collapses under it (Table~\ref{tab:app-fixedref}), and our read-age
results (Section~\ref{sec:r-age}) suggest that the other drift-sensitive baselines would too. We instead use a
reference column that ages with the weights (Appendix~\ref{app:eff-wbits}). Regular blocks cost nothing where
plain blocks suffice and prevent collapse where they fail: with a reference column both perform alike at 8-bit
ADCs, but with a fixed reference plain blocks collapse to 0.0\% while regular blocks hold at 82.7\%
(Table~\ref{tab:app-fixedref}).

\begin{table}[H]
\centering\footnotesize\setlength{\tabcolsep}{4pt}
\caption{ViT-B/16 top-1 accuracy (\%) on CMO ReRAM with 8-bit ADCs five days after programming, with our method
using regular or plain output blocks, on 4,898 evaluation images with one seed.}
\label{tab:app-fixedref}
\begin{tabular}{@{}lrrr@{}}
\toprule
Reference & Regular blocks & Plain blocks & Unmodified \\
\midrule
Reference column & 82.9 & 82.7 & 24.8 \\
Fixed reference & 82.7 & 0.0 & 0.1 \\
\bottomrule
\end{tabular}
\end{table}

\section{Drift compensation granularity}\label{app:drift-gain}

Global drift compensation \cite{Joshi2020} usually estimates one gain per array. Our method needs only one gain
per layer, folded into the inverse output rotation: the rotations make the arrays of a layer statistically
alike, so they drift alike and per-array compensation would assign them nearly the same gain
(Table~\ref{tab:drift-gain}). With our method, these gains differ by a median below 1\%, against up to 22\%
without it. One gain per layer therefore leaves our local output NMSE unchanged.

\begin{table}[H]
\centering\footnotesize\setlength{\tabcolsep}{4pt}
\caption{Gain spread (\%) of per-array compensation, PCM with 4-bit ADCs at one day: the relative difference
between the gains it assigns to the arrays of one layer, as median / maximum over layers.}
\label{tab:drift-gain}
\begin{tabular}{@{}lrr@{}}
\toprule
Model & \textit{Unmodified} & \textbf{Ours} \\
\midrule
ViT & \textit{5.7 / 21} & \textbf{0.66 / 2.8} \\
ResNet & \textit{2.0 / 13} & \textbf{0.21 / 2.2} \\
Qwen & \textit{7.9 / 22} & \textbf{0.92 / 2.4} \\
\bottomrule
\end{tabular}
\end{table}

\section{Device noise strength}
\label{app:noise}

\begin{figure}[t]
\centering
\includegraphics[width=\columnwidth]{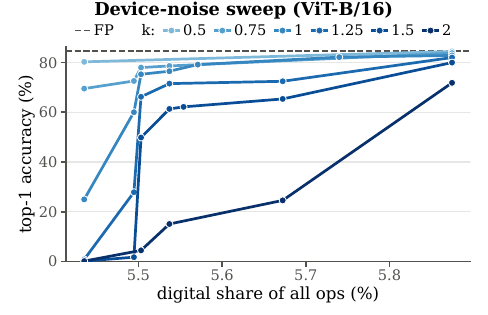}
\caption{ViT-B/16 accuracy against the digital share of all analog and digital operations, on ReRAM with
8-bit ADCs at five days, with programming, read and drift noise scaled by $k$. The dashed line marks
floating-point accuracy.}
\label{fig:s52-noise}
\end{figure}

Figure~\ref{fig:s52-noise} repeats the cost and accuracy trade-off of Section~\ref{sec:r-cost} for
ViT-B/16 with programming, read and drift noise scaled by $k$. At half the default noise, the
unmodified model is already close to floating-point accuracy, and only the full method improves on it. As noise
increases, more of the method is needed to reach the same accuracy, and at twice the default noise most of the
gain comes from the most complete configurations.

\section{IR drop}\label{app:irdrop}

IR drop is negligible in our tests. With the AIHWKit IR-drop model at its default wire resistance and a
maximum conductance of 25\,$\mu$S, the worst rotated ViT-B/16 or Qwen3-1.7B projection has an IR-drop RMS
error of about 4\% of its output RMS, while the step of a 4-bit ADC under our range rule is about half the
output RMS. End to end, the accuracy of our method on ViT-B/16 changes by less than 0.4 points on PCM, even at
twice the default wire resistance (Table~\ref{tab:irdrop}).

\begin{table}[H]
\centering\footnotesize\setlength{\tabcolsep}{2.5pt}
\caption{ViT-B/16 top-1 accuracy (\%) on PCM at one day, as ours/unmodified, without IR drop and with IR
drop at 1, 1.5 and 2 times the default wire resistance.}
\label{tab:irdrop}
\begin{tabular}{@{}lcccc@{}}
\toprule
ADC & IR off & $1\times$ & $1.5\times$ & $2\times$ \\
\midrule
4 & 80.2/4.0 & 80.0/3.9 & 80.1/3.7 & 79.8/3.7 \\
6 & 82.2/44.1 & 82.3/45.0 & 82.5/44.6 & 82.3/45.2 \\
8 & 82.5/44.4 & 82.5/44.3 & 82.5/44.6 & 82.7/44.5 \\
\bottomrule
\end{tabular}
\end{table}

\section{Hardware-aware training comparison setup}\label{app:hwa-setup}
\textbf{ResNet comparison.} For ResNet-18/50, we follow the direct-mapping setting of
\citet{Rasch2023}: PCM programming, drift and read noise or the ReRAM model of
\citet{Wan2022}, 8-bit ADCs with bounds $\pm10$, a static 8-bit DAC, per-MVM read and output noise,
and IR drop. We evaluate one day after programming, with probe-based post-ADC global drift
compensation on PCM and none on ReRAM. All 50,000 ImageNet validation images are evaluated, and
calibration uses training images. Analog weights are not trained, while digital input scales,
per-column output scales, and BatchNorm parameters are fitted as in \citet{Rasch2023}.

\textbf{AFM comparison.} We reproduce the AFM Hermes PCM programming-noise setting
of \citet{Buchel2025AFM}. At its nominal noise ($k=1.00$), our noise-only model scores 77.9\% on Phi-3-mini
and 40.7\% on Llama-3.2-1B, above the published 76.5\% and 38.5\%. Our reproduction is therefore milder
than theirs, and comparing at nominal noise would favor our method. The only way we could match their setting
was to increase the programming-noise standard deviation by 10\% ($k=1.10$), which brings the noise-only scores
to 76.5\% and 38.2\%. The main text uses this matched setting. Table~\ref{tab:afm-comp} includes our
measurements at both noise levels to show the adjustment: our method scores 79.3\% and 44.5\% at nominal
noise, and 78.4\% and 42.8\% at $k=1.10$.
Checkpoints: \texttt{microsoft/Phi-3-mini-4k-instruct}, revision\\
{\small\texttt{ff07dc01615f8113924aed013115ab2abd32115b}}, and \texttt{meta-llama/Llama-3.2-1B-Instruct}, revision\\
{\small\texttt{9213176726f574b556790deb65791e0c5aa438b6}}. Phi-3-mini results outside this comparison use the latest revision,\\
{\small\texttt{f39ac1d28e925b323eae81227eaba4464caced4e}}.

SI8-W16h-O8 uses 8-bit output quantization with per-layer calibrated ranges. Scores
average MMLU, HellaSwag, ARC-Easy, and ARC-Challenge using the ARC answer-key convention of \citet{Buchel2025AFM}.

\section{QuaRot comparison}
\label{app:quarot}

Table~\ref{tab:quarot} compares QuaRot \cite{Ashkboos2024QuaRot}, applied to the arrays directly, with the
unmodified model and our method. QuaRot was designed for digital quantizers. It does not center inputs, and its
randomized plain Hadamard blocks behave like the DHP blocks of Figure~\ref{fig:plain-vs-regular}, which raise
4-bit ADC error above the unrotated computation. Appendix~\ref{app:loo} removes our components one at a time.

\begin{table}[H]
\centering\footnotesize\setlength{\tabcolsep}{2.5pt}
\caption{LLM mean accuracy (\%) of the unmodified model (U), QuaRot (Q) and our method (O) with 4-bit ADCs,
one day after programming. Phi models are scored on 512 questions per task for all three methods.}
\label{tab:quarot}
\begin{tabular}{@{}llrrrr@{}}
\toprule
Model & Device & U & Q & O & FP \\
\midrule
Llama-3.2-1B & PCM & 24.6 & 26.1 & \textbf{41.0} & 50.1 \\
 & ReRAM & 24.2 & 26.5 & \textbf{42.0} &  \\
Qwen3-1.7B & PCM & 23.9 & 31.3 & \textbf{47.2} & 59.0 \\
 & ReRAM & 23.7 & 29.6 & \textbf{48.2} &  \\
Phi-3-mini & PCM & 25.7 & 27.3 & \textbf{60.2} & 68.4 \\
 & ReRAM & 24.9 & 26.5 & \textbf{61.2} &  \\
Qwen3-8B & PCM & 23.9 & 47.7 & \textbf{62.3} & 70.7 \\
 & ReRAM & 24.4 & 52.4 & \textbf{63.1} &  \\
Phi-4 & PCM & 58.1 & 25.6 & \textbf{69.4} & 74.4 \\
 & ReRAM & 31.5 & 26.6 & \textbf{70.9} &  \\
\bottomrule
\end{tabular}
\end{table}

\begingroup
\setlength{\intextsep}{6pt}

\section{Effective weight precision}\label{app:eff-wbits}
\begin{table}[H]
\centering
\caption{Effective precision $b_{\mathrm{eff}}$ of continuous ViT-B/16 conductances versus read age.
Parentheses give effective precision computed from the worst conditional RMS error over the
conductance grid. GDC denotes global drift compensation.}
\label{tab:eff-wbits}
\begingroup
\fontsize{7.5}{9}\selectfont
\setlength{\tabcolsep}{2pt}
\begin{tabular*}{\linewidth}{@{\extracolsep{\fill}}lcccc@{}}
\toprule
Device / configuration & 1 s & 1 h & 1 day & 5 days \\
\midrule
\textit{PCM, pair + GDC} & \textit{\shortstack{4.62\\(3.32)}} & \textit{\shortstack{4.44\\(2.55)}} & \textit{\shortstack{4.27\\(2.06)}} & \textit{\shortstack{4.17\\(1.86)}} \\
\cmidrule(lr){1-5}
\textit{\shortstack[l]{ReRAM 0.2\%\\fixed ref.}} & \textit{\shortstack{5.67\\(5.60)}} & \textit{\shortstack{4.41\\(4.39)}} & \textit{\shortstack{4.06\\(4.04)}} & \textit{\shortstack{3.91\\(3.89)}} \\
\textit{\shortstack[l]{ReRAM 2\%\\fixed ref.}} & \textit{\shortstack{4.99\\(4.40)}} & \textit{\shortstack{4.22\\(3.95)}} & \textit{\shortstack{3.93\\(3.74)}} & \textit{\shortstack{3.80\\(3.63)}} \\
\cmidrule(lr){1-5}
\textit{\shortstack[l]{ReRAM 0.2\%\\fixed ref.} + GDC} & \textit{\shortstack{5.69\\(4.18)}} & \textit{\shortstack{4.48\\(2.28)}} & \textit{\shortstack{4.17\\(1.84)}} & \textit{\shortstack{4.04\\(1.66)}} \\
\textit{\shortstack[l]{ReRAM 2\%\\fixed ref.} + GDC} & \textit{\shortstack{5.00\\(4.07)}} & \textit{\shortstack{4.30\\(2.38)}} & \textit{\shortstack{4.04\\(1.92)}} & \textit{\shortstack{3.93\\(1.73)}} \\
\cmidrule(lr){1-5}
\textit{\shortstack[l]{ReRAM 0.2\%\\physical ref.}} & \textit{\shortstack{5.17\\(5.14)}} & \textit{\shortstack{4.36\\(4.32)}} & \textit{\shortstack{4.14\\(4.00)}} & \textit{\shortstack{4.03\\(3.86)}} \\
\textit{\shortstack[l]{ReRAM 2\%\\physical ref.}} & \textit{\shortstack{4.50\\(4.15)}} & \textit{\shortstack{4.07\\(3.85)}} & \textit{\shortstack{3.91\\(3.73)}} & \textit{\shortstack{3.83\\(3.67)}} \\
\bottomrule
\end{tabular*}
\endgroup
\end{table}

\smallskip
\noindent\textbf{Definition.}
$b_{\mathrm{eff}}=\log_2[2w_{\max}/(\sqrt{12}\,\sigma_w)]$, with
$\sigma_w^2=\mathbb{E}[(\widetilde w-w)^2]$ including bias. Estimates cover all ViT transformer
linear weights with per-array full-range mapping. GDC uses the per-array mean-absolute-weight ratio.

\smallskip
\noindent\textbf{Reference.} AIHWKit subtracts a fixed, ideal reference value by default. Our main ReRAM setting
instead subtracts a programmed reference column that ages with the weights, so the reference tracks their
drift. This removes much of the shared drift error that a fixed reference would add, which makes the comparison
fairer to the baselines. The physical-reference, 0.2\%-acceptance row without GDC matches the main ReRAM
setting, and the fixed-reference rows show the default for comparison.

\smallskip
\noindent\textbf{ReRAM parameters.}
We use the 0.2\% programming-acceptance variant, using the fits from \citet{Falcone2025}.
\endgroup

\end{document}